\documentclass{article}

\usepackage[utf8]{inputenc} 
\usepackage[T1]{fontenc}    
\usepackage{hyperref}       
\usepackage{url}            
\usepackage{booktabs}       
\usepackage{amsfonts}       
\usepackage{nicefrac}       
\usepackage{microtype}      
\usepackage{lipsum}
\usepackage{natbib}
\usepackage{fancyhdr}       
\usepackage{graphicx}       
\usepackage{wrapfig}
\usepackage{amsmath}
\usepackage{subcaption} 
\usepackage{caption}
\usepackage{tabularx}
\usepackage{float}
\usepackage{authblk} 
\usepackage{setspace} 

\begin{document}

\title{\LARGE \bf HDR Image Reconstruction using an
 Unsupervised Fusion Model } 
\author[1]{Kumbha Nagaswetha}

\affil[1]{Department of Electrical Communication Engineering, \newline Indian Institute of Science, Bengaluru, India, \texttt{nagaswetha.k@alum.iisc.ac.in}}
\date{} 

\maketitle

\begin{abstract}
    High Dynamic Range (HDR) imaging aims to reproduce the wide range of brightness levels present in natural scenes, which the human visual system can perceive but conventional digital cameras often fail to capture due to their limited dynamic range. To address this limitation, we propose a deep learning-based multi-exposure fusion approach for HDR image generation. The method takes a set of differently exposed Low Dynamic Range (LDR) images, typically an underexposed and an overexposed image, and learns to fuse their complementary information using a convolutional neural network (CNN). The underexposed image preserves details in bright regions, while the overexposed image retains information in dark regions; the network effectively combines these to reconstruct a high-quality HDR output. The model is trained in an unsupervised manner, without relying on ground-truth HDR images, making it practical for real-world applications where such data is unavailable. We evaluate our results using the Multi-Exposure Fusion Structural Similarity Index Measure (MEF-SSIM) and demonstrate that our approach achieves superior visual quality compared to existing fusion methods. A customized loss function is further introduced to improve reconstruction fidelity and optimize model performance.
\end{abstract}

\section{Introduction}

The range of brightness in real-world scenes often exceeds what conventional digital cameras can capture. While the human visual system can adapt to extreme luminance variations, typical camera sensors are limited in their dynamic range, resulting in overexposed or underexposed regions when photographing high-contrast scenes \cite{HAL, hdr}. Low Dynamic Range (LDR) images captured by a single exposure often fail to preserve details in shadows or highlights, losing critical scene information \cite{hdrvsldr}. For example, underexposed images may preserve bright regions such as the sun but lose shadow details, while overexposed images reveal shadows at the cost of saturating bright regions \cite{dataset}.

An image is formally represented as a two-dimensional function $f(x,y)$, where $x$ and $y$ are spatial coordinates and the discrete amplitude values correspond to pixel intensities. Pixel values typically range from 0 (black) to 255 (white) \cite{image-capture}. The brightness captured depends on exposure time, aperture, and sensor sensitivity. Longer exposure times increase light intake but may cause overexposure, while shorter exposures can underexpose darker regions. For dynamic scenes, longer exposures can introduce motion blur or ghosting, whereas short exposures can freeze motion \cite{blur}. In this work, we focus on static scenes where such motion artifacts are absent.

The dynamic range (DR) of an image quantifies the ratio between the maximum and minimum measurable pixel intensities:
\begin{equation}
    DR = \log_{10} \frac{I_\text{max}}{I_\text{min}}
\end{equation}
where $I_\text{max}$ and $I_\text{min}$ represent the maximum and minimum pixel intensities, respectively \cite{hdr}. While natural scenes can have a DR exceeding 224, standard digital cameras typically capture scenes with DRs between 28 and 212. This discrepancy highlights the need for techniques that extend the effective dynamic range of captured images.

High Dynamic Range (HDR) imaging addresses these limitations by reconstructing images that preserve details across the full luminance spectrum \cite{mertens, RC, patches}. HDR images combine information from multiple exposures or enhance single LDR images to produce outputs that more closely resemble human perception \cite{hdrvsldr}. In multi-exposure HDR techniques, complementary details are extracted from several LDR images with different exposures and fused into a single high-fidelity image \cite{bruce, lee, RC}. Challenges include motion between exposures (causing ghosting), computational cost, and real-time constraints. Single-image HDR methods generate multiple pseudo-exposures internally and combine them using tone-mapping or filtering \cite{sef}, but these approaches may alter edges, reduce resolution, or introduce artifacts.

Recently, deep learning approaches have demonstrated strong performance in HDR reconstruction due to their ability to model complex nonlinear mappings \cite{cnn}. Supervised methods train convolutional neural networks (CNNs) on paired LDR-HDR datasets to learn exposure fusion mappings \cite{kalantari, li}, while unsupervised methods employ loss functions and perceptual metrics to optimize HDR reconstruction without requiring ground-truth HDR images \cite{prabhakar, qi, hdrcnn, loss}. These methods can generate high-quality HDR images with fewer artifacts and better preservation of structural details, though they require substantial training data and computational resources.

In this work, we review and compare different computational HDR methods, including multi-exposure fusion, single-image reconstruction, and deep learning-based techniques. Our focus is on enhancing dynamic range, preserving structural details, and improving visual quality using methods applicable to conventional camera images, without hardware modifications. By addressing the limitations of LDR imaging, these approaches enable more faithful representations of real-world scenes in photography, television, biomedical imaging, and surveillance applications.

\section{Overview of Methods}

High Dynamic Range (HDR) imaging techniques aim to capture and represent the full range of luminance present in real-world scenes, which often exceed the capability of standard imaging sensors. Several approaches have been developed to obtain HDR images depending on the input data and target application. Broadly, these methods can be classified into three categories: (i) techniques based on multiple Standard Dynamic Range (SDR) images, (ii) single image methods, and (iii) deep learning based approaches \cite{methods}.

\subsection{Using Multiple SDR Images}

The most common and traditional approach involves capturing multiple SDR images of the same scene with different exposure times and combining them to form a single HDR image. Each exposure captures specific luminance information underexposed images preserve details in bright areas, while overexposed ones retain information in dark regions. A fusion process or tone mapping algorithm then integrates these complementary features to produce an HDR image that maintains well-exposed details across the full intensity range \cite{bruce,lee,RC}.

Although this method generates visually high quality HDR images, it has certain limitations. The input images must be perfectly aligned, which becomes challenging in scenes containing motion or when handheld cameras are used. Any slight movement between exposures can introduce ghosting artifacts. Moreover, acquiring and processing multiple images increases computational complexity and limits real-time applicability.

\subsection{Using a Single SDR Image}

Single-image HDR reconstruction methods attempt to synthesize HDR-like content from one SDR input. These methods often apply enhancement operations such as local contrast stretching, tone mapping, or filtering. The key idea is to simulate a range of virtual exposures internally and merge them to reconstruct the dynamic range \cite{sef}.

Such approaches are computationally efficient and suitable for real-time use since they do not rely on multiple captures. However, they often fail to accurately recover information in saturated or clipped regions, leading to visual artifacts and reduced fidelity compared to true HDR imaging. The reconstructed images may lose edge sharpness and structural consistency, especially in challenging lighting conditions.

\subsection{Using Deep Learning Methods}

Deep learning has emerged as a powerful alternative to traditional HDR generation techniques. Convolutional Neural Networks (CNNs) and other architectures can effectively learn the mapping between SDR and HDR domains by exploiting large datasets. Depending on data availability, these models operate in either supervised or unsupervised settings. Supervised approaches use paired SDR–HDR images for training \cite{kalantari,li}, while unsupervised ones learn from unpaired or synthetic data using consistency and perceptual loss functions \cite{prabhakar,qi}.

Deep learning models have demonstrated strong generalization ability and robustness to motion, noise, and exposure variation. They can infer HDR content from even a single SDR image, producing visually realistic results. However, these methods demand significant computational resources and extensive training data, which limits their scalability in resource-constrained environments.

\subsection{Multi-Exposure Image Fusion Based on Adaptive Weights}

Among the multiple-exposure methods, one effective approach to HDR generation is multi-exposure image fusion (MEF), where differently exposed images are combined using adaptively computed weights. The central idea is to assign higher weights to image regions that are well-exposed and contain more structural information, ensuring that the fused HDR image retains the best details from all inputs \cite{lee}.

In the proposed adaptive-weight method, two complementary weighting functions are designed. The first weight is based on well-exposedness\cite{mertens}, giving importance to pixels with intensity values near the mid-range of 0.5 after normalization. To account for image brightness variations, the weight distribution is made adaptive by incorporating the mean intensity of each exposure, thereby assigning higher weights to informative regions in both underexposed and overexposed images.

The second weighting scheme is gradient-based, motivated by the human visual system’s sensitivity to local contrast and intensity variation. Gradients are computed from cumulative histograms of input images regions with large gradients correspond to areas containing meaningful structural information. By inversely weighting these gradients, the method emphasizes well-exposed regions and suppresses saturated areas.

The final composite weight is obtained by multiplying the two weight maps and normalizing them across all input images. The fused HDR image $I_{fused}$ is computed as a weighted sum of the input exposures:
\begin{equation}
    I_{fused}(x, y) = \sum_{n=1}^{N} W_n(x, y) \, I_n(x, y)    
\end{equation}

where ${W_n(x, y)}$ is the normalized combined weight for the $n^{th}$ exposure.

\subsubsection{Experimental Results}

\begin{figure}
\centering
    \begin{subfigure}[b]{0.3\textwidth}
        \centering
        \includegraphics[width=\textwidth]{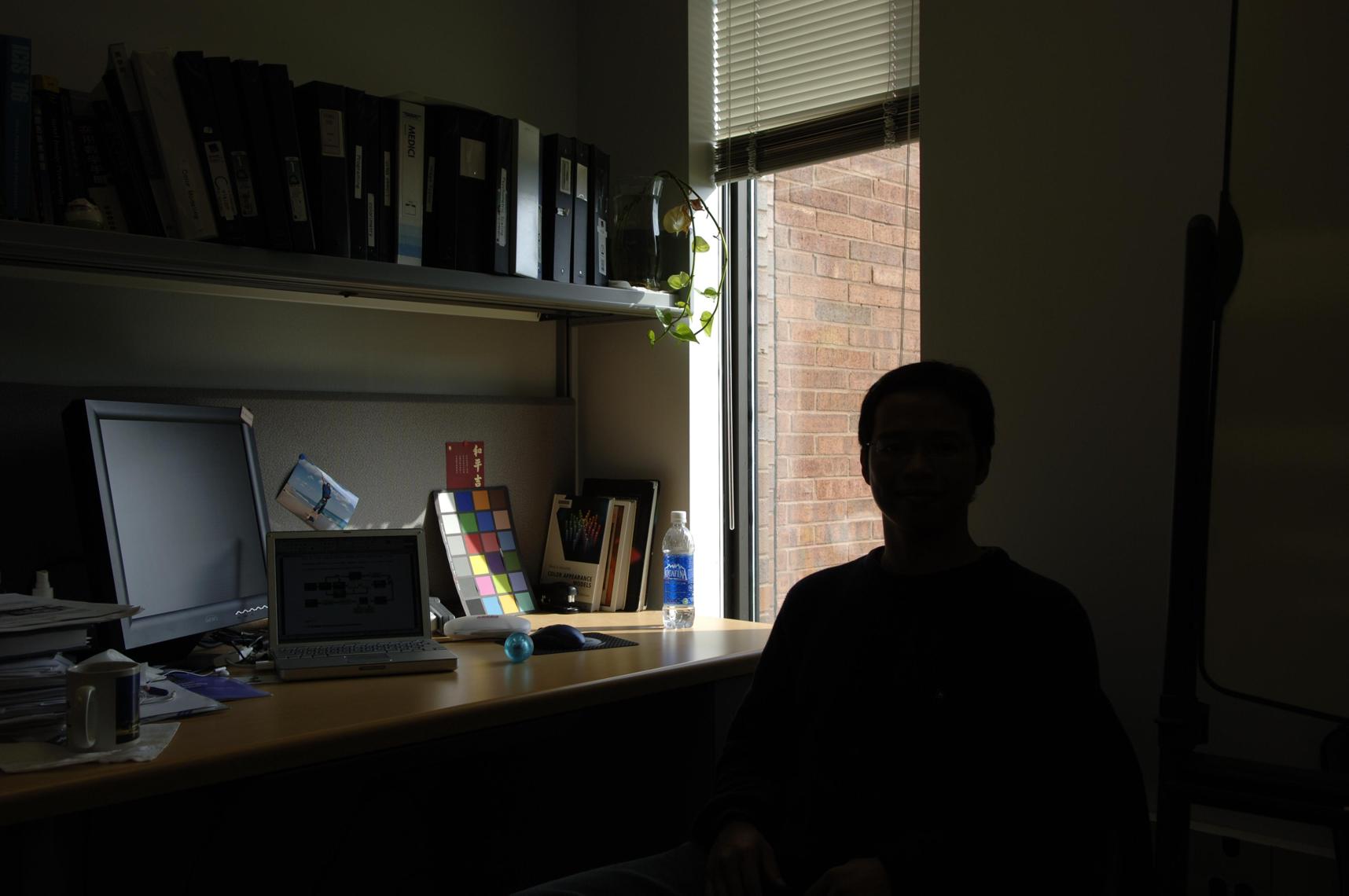}
        \caption{Underexposed input}
        \label{fig:under}
    \end{subfigure}
\quad
\quad
    \begin{subfigure}[b]{0.3\textwidth}
        \centering
        \includegraphics[width=\textwidth]{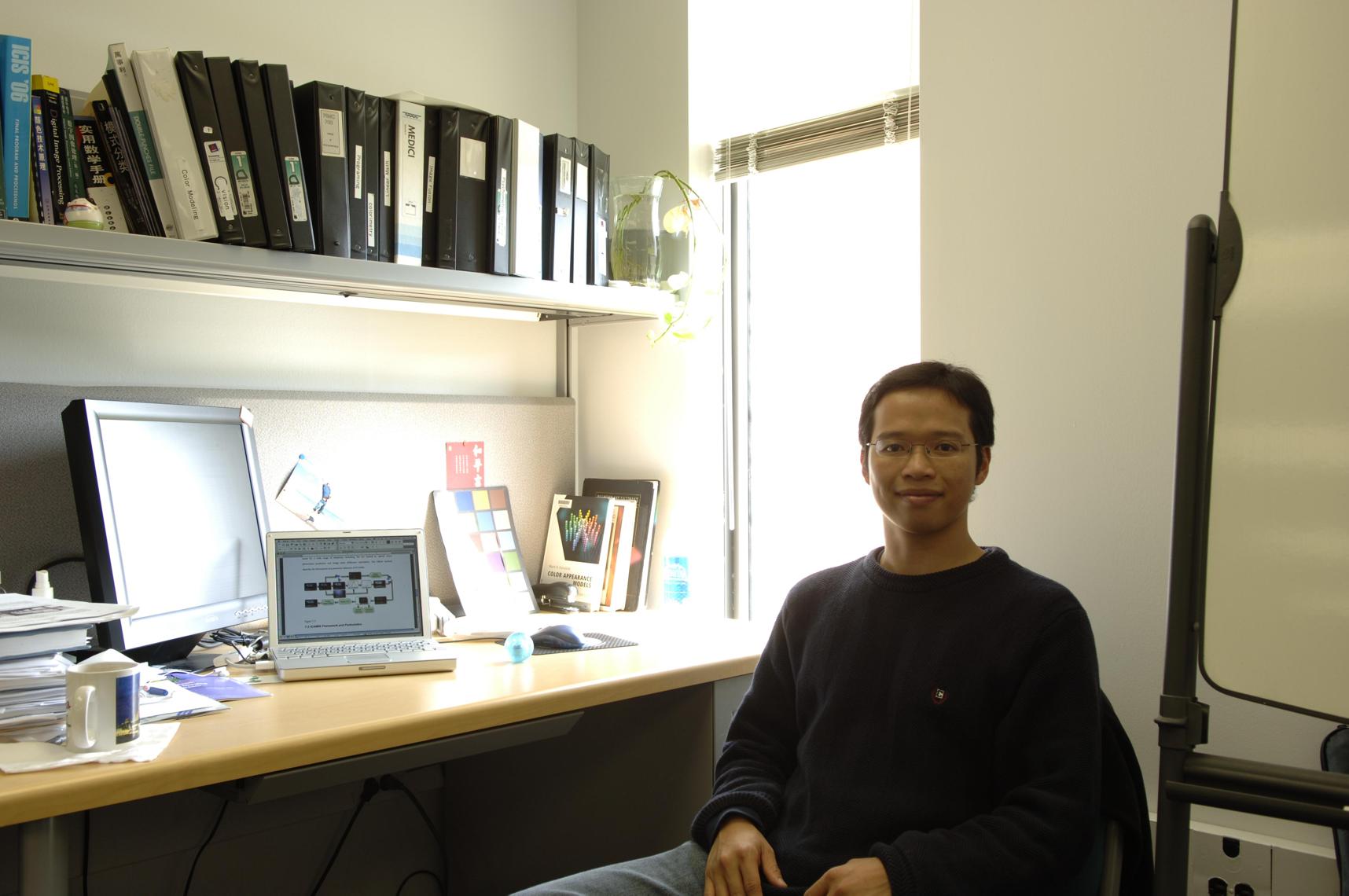}
        \caption{Overexposed input}
        \label{fig:over}
    \end{subfigure}
\quad
    \begin{subfigure}[b]{0.3\textwidth}
        \centering
        \includegraphics[width=\textwidth]{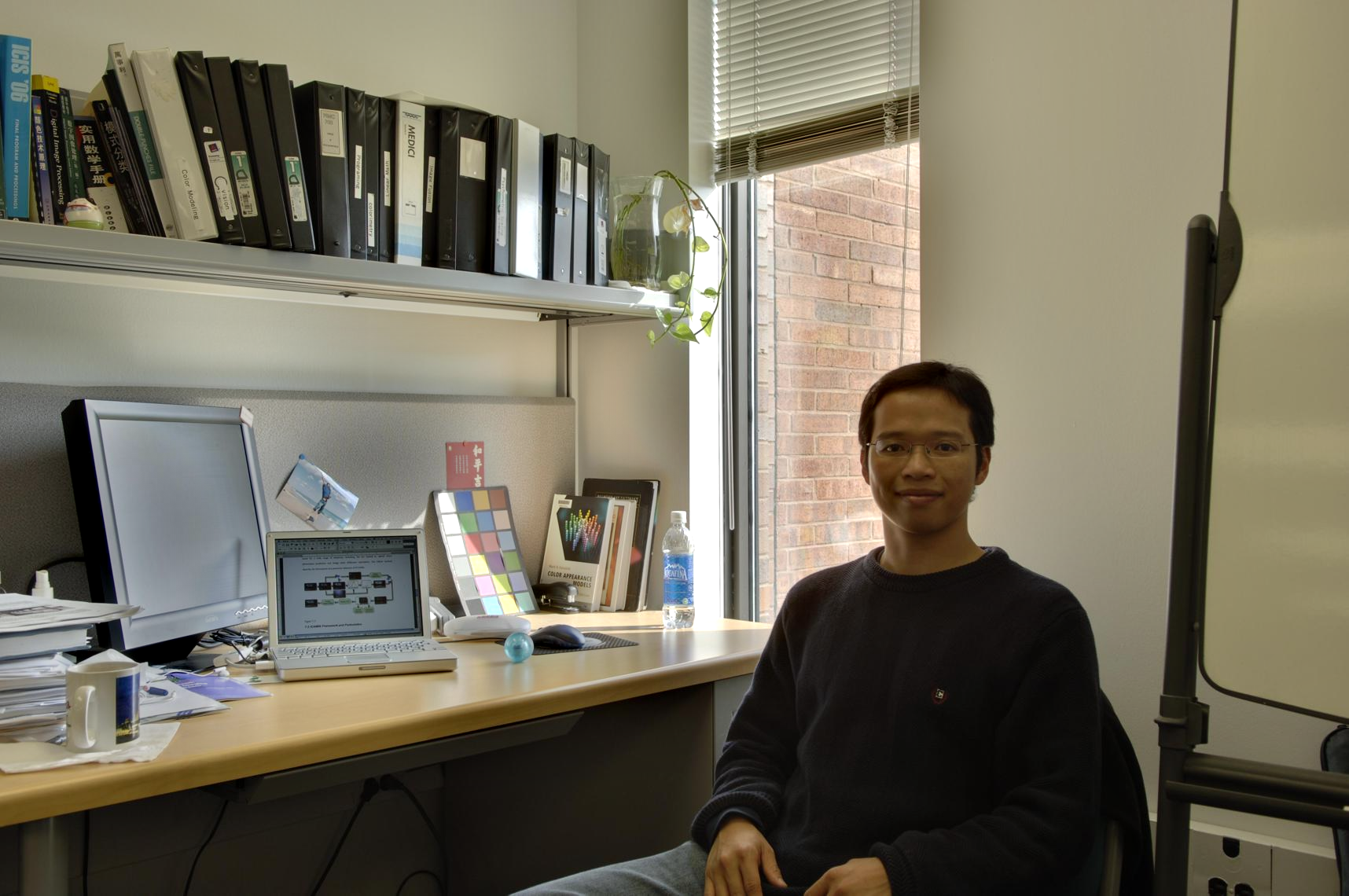}
        \caption{HDR output (both weights)}
        \label{fig:both}
    \end{subfigure}
\quad
    \begin{subfigure}[b]{0.3\textwidth}
        \centering
        \includegraphics[width=\textwidth]{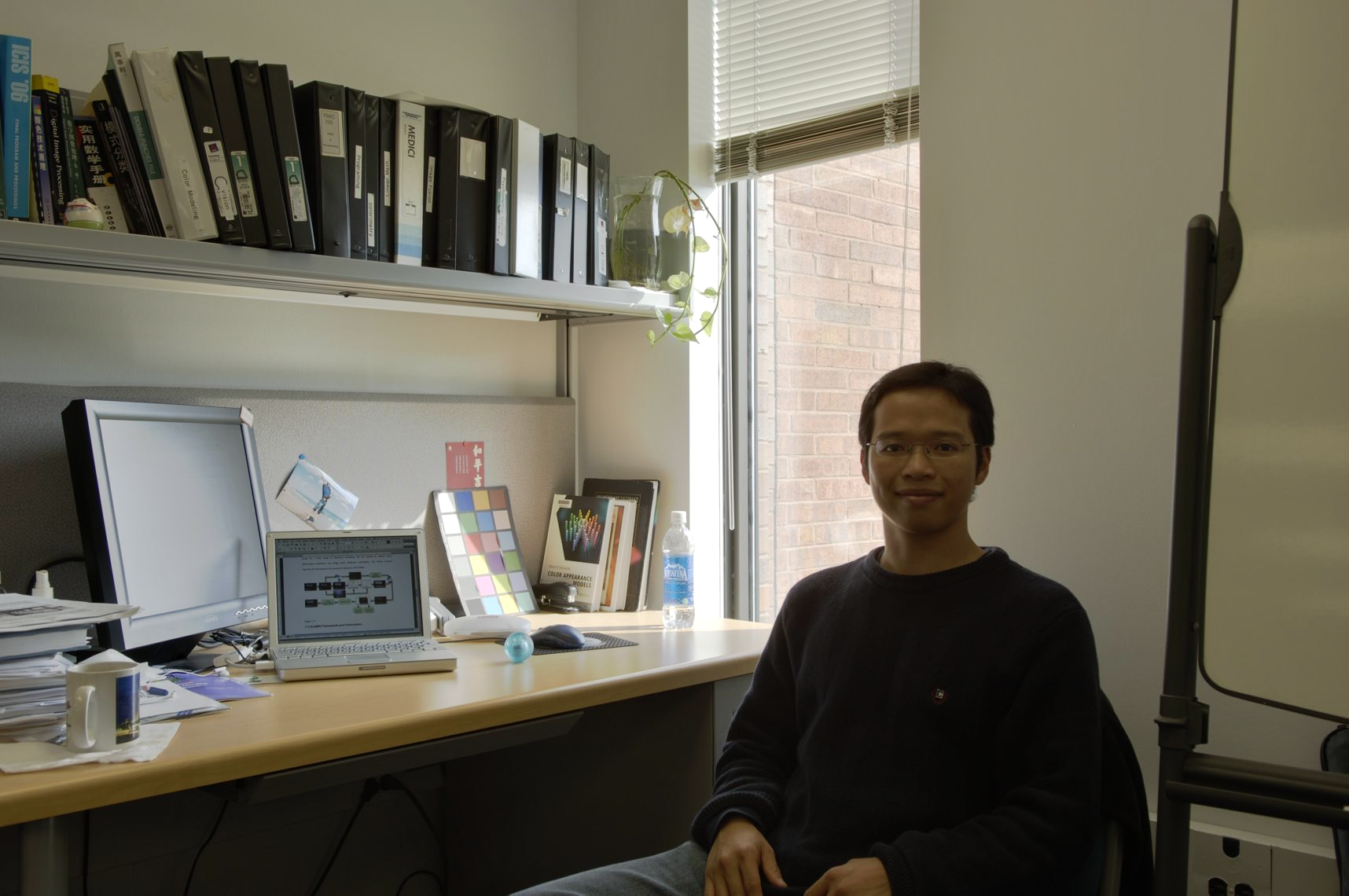}
        \caption{HDR output (Weight 1 only)}
        \label{fig:w1}
    \end{subfigure}
\quad
    \begin{subfigure}[b]{0.3\textwidth}
        \centering
        \includegraphics[width=\textwidth]{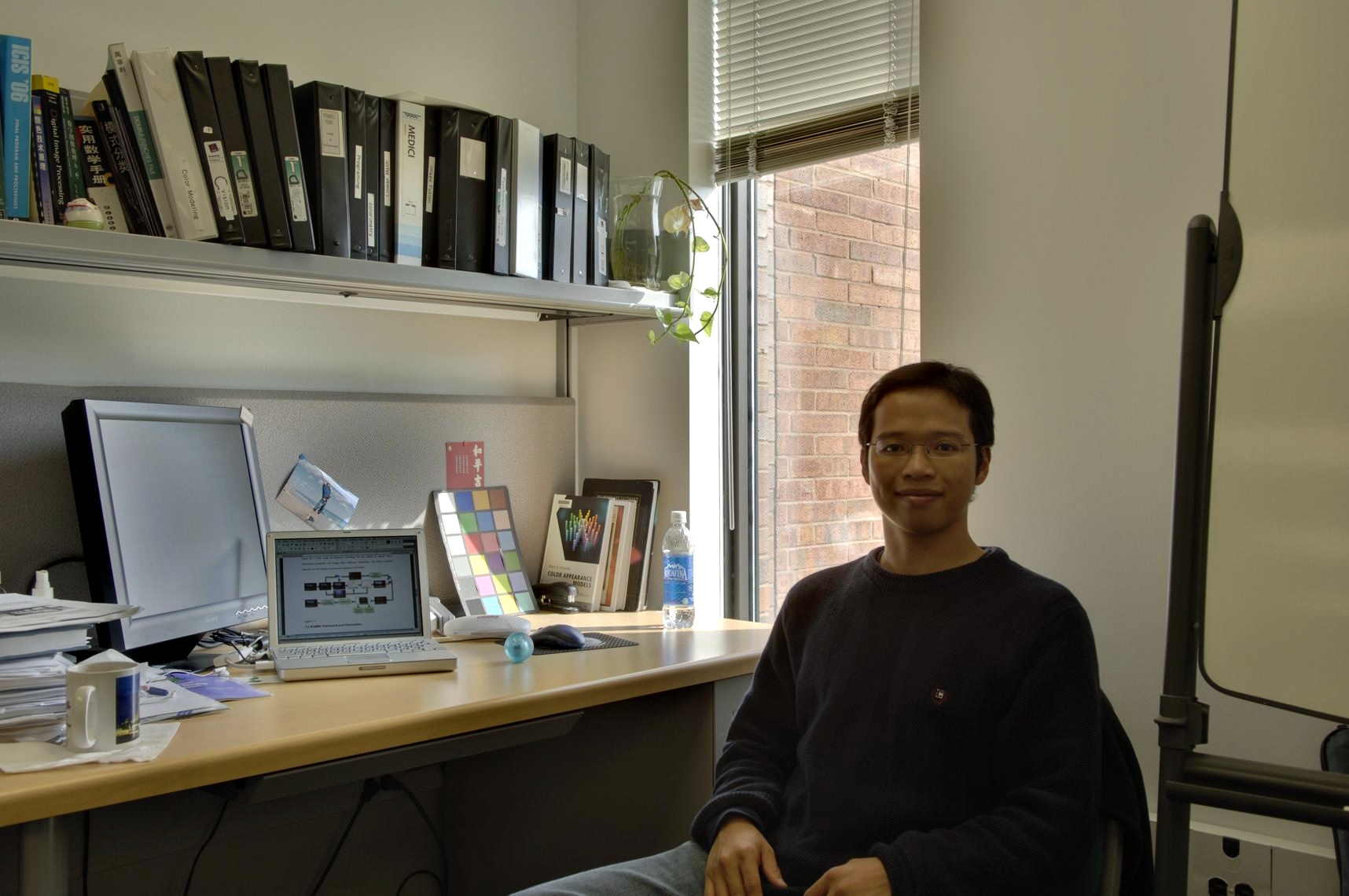}
        \caption{HDR output (Weight 2 only)}
        \label{fig:w2}
    \end{subfigure}
    \caption{Comparison of MEF outputs: combined weights produce balanced HDR images with improved detail preservation.}
    \label{fig:mef_results}
\end{figure}

The MEF method was evaluated on multiple image sequences, and the Multi-Exposure Fusion Structural Similarity Index (MEF-SSIM) \cite{mefssim} was used to quantify visual fidelity. Figure~\ref{fig:mef_results} illustrates the HDR outputs when using individual weights versus the combined adaptive weights.

When only Weight 1 is applied, bright areas such as brick regions appear overexposed, whereas applying only Weight 2 improves structural detail in mid-tones but introduces dark artifacts in other regions. Using both weights simultaneously mitigates these issues, producing visually balanced HDR images without overexposed or underexposed artifacts.

Table~\ref{tab:mef_scores} reports the MEF-SSIM scores for five test sequences. The proposed adaptive-weight fusion method achieves consistently high fidelity, with an average MEF-SSIM of 0.9207, demonstrating improved preservation of luminance and structural details compared to single-weight schemes.

\begin{table}[ht!]
\centering
\begin{tabularx}{0.7\textwidth} { 
  | >{\raggedright\arraybackslash}X 
  | >{\centering\arraybackslash}X | }
 \hline
 \textbf{Test Images} & \textbf{MEF-SSIM score} \\
 \hline
 Balloons         & 0.8778 \\
 \hline
 Belgium House    & 0.9715 \\
 \hline
 House            & 0.9455 \\
 \hline
 Light House      & 0.8627 \\
 \hline
 Madison Capital  & 0.9461 \\
 \hline
 \textbf{Average} & \textbf{0.9207} \\
 \hline
\end{tabularx}
\caption{MEF-SSIM scores for the proposed adaptive-weight fusion method.}
\label{tab:mef_scores}
\end{table}

These results confirm that the adaptive-weight MEF approach effectively combines complementary information from multiple exposures, leading to HDR images that preserve details in both bright and dark regions while minimizing artifacts.

\section{HDR Image Reconstruction Using an Unsupervised Learning Model}
Traditional HDR reconstruction methods, such as multi-exposure fusion (MEF), rely on combining differently exposed images using handcrafted weights. While effective, these approaches often produce artifacts in extreme lighting conditions and require careful tuning \cite{lee,mertens}. Deep learning-based HDR generation can represent complex image mappings and reconstruct missing information, outperforming classical methods \cite{hdrcnn}. However, supervised HDR networks require ground truth HDR images, which are difficult to obtain. To address these challenges, we propose an unsupervised CNN-based HDR reconstruction model that learns to predict per-pixel weight maps from multi-exposure inputs.

\subsection{Model Architecture}
The proposed network follows an encoder-decoder architecture, inspired by VGG16 \cite{vgg16}, to extract features from multi-exposure inputs and reconstruct weight maps for HDR fusion. Let $I_n$ denote the $n^{th}$ input image and $W_n$ the corresponding predicted weight map. The fused HDR image is obtained as:

\begin{equation}
I_{\text{fused}}(c) = \sum_{n=1}^{N} W_n \cdot I_n(c), \quad c \in \{R, G, B\}
\end{equation}

\begin{figure}[!ht]
\centering
\includegraphics[height=3cm,width=1.2\textwidth]{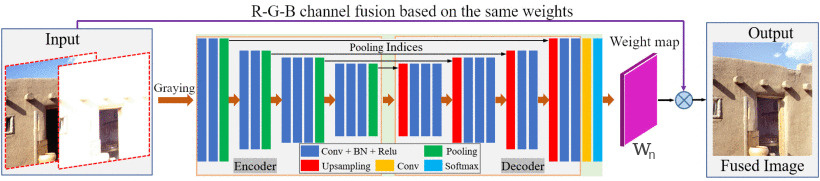}
\caption{Overall CNN architecture for unsupervised HDR reconstruction. The encoder extracts feature maps, and the decoder reconstructs the weight maps $W_n$ used for HDR fusion.}
\end{figure}

\subsubsection{Encoder}
The encoder takes two input images (under and over-exposed), converted to grayscale to reduce computational cost. Each input is processed through a series of convolutional layers, batch normalization, ReLU activation, and max-pooling. This module contains 10 convolutional layers with filter sizes: 64, 64, 128, 128, 256, 256, 256, 512, 512, 512.

\subsubsection{Decoder}
The decoder mirrors the encoder structure, performing upsampling followed by convolution, batch normalization, and ReLU. The filter sizes are reversed: 512, 512, 512, 256, 256, 256, 128, 128, 64, 64. The output is processed through a softmax layer to produce the weight map for each input image.

\begin{figure}[!ht]
\centering
\subfloat[Encoder \label{fig:encoder}]{\includegraphics[height=15cm,width=0.4\textwidth]{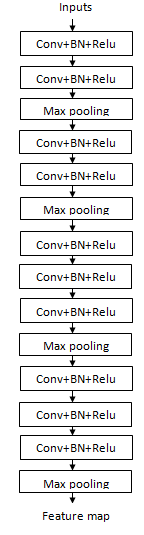}}
\qquad
\subfloat[Decoder \label{fig:decoder}]{\includegraphics[height=15cm,width=0.4\textwidth]{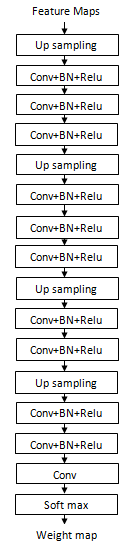}}
\caption{Encoder and Decoder sub-modules of the network. Encoder extracts features, and decoder reconstructs weight maps used for HDR fusion.}
\end{figure}

\subsection{Proposed Loss Function}
Training is unsupervised, without ground truth HDR images. Standard SSIM \cite{ssim} cannot directly handle multi-exposure inputs. Therefore, we use a weighted SSIM loss function \cite{loss}, which combines contributions from both under and over-exposed images:

\begin{multline*}
loss = 1 - \frac{1}{15}\sum\limits _{c \in \lbrace R,G,B\rbrace } {\sum \limits _w {\left[{{\gamma _w} \cdot {ssim}\left({{{{ I}}_{un\exp }}(c),{{{ I}}_f}(c);w} \right)} \right.}} \\ 
\left. { + \left({1 - {\gamma _w}} \right) \cdot {ssim}\left({{{{I}}_{ov\exp }}(c),{{{ I}}_f}(c);w} \right)} \right]
\end{multline*}

where $\gamma_w$ is the local variance-based weight:

\begin{equation}
\gamma_w = \frac{g(\sigma^2_{w_{I_{\text{unexp}}}})}{g(\sigma^2_{w_{I_{\text{unexp}}}}) + g(\sigma^2_{w_{I_{\text{ovexp}}}})}, \quad g(x) = \max(x, 10^{-4})
\end{equation}

The crucial factor influencing the weighted SSIM loss function in Equation~(4) is $\gamma$. It determines how much each exposure contributes to the final HDR estimate. Ideally, $\gamma$ should be sensitive to well-exposed regions, differentiable for backpropagation, and robust against extreme luminance variations. To explore this, we examine multiple formulations of $\gamma$ derived from three perceptual attributes variance, gradient, and well-exposedness \cite{mertens}. These represent distinct yet complementary cues for identifying informative image regions.

The equations for combining these parameters are shown below. The variance and gradient are combined as:

\begin{equation}
    \gamma_{un\exp} =\frac{\sigma^2_{un \exp} \times Grad_{un \exp}}{\sqrt{(\sigma_{un \exp}^{2})^{2}+ Grad_{un \exp}^{2}}} 
\end{equation}
where $\sigma ^{2}$ denotes local intensity variance and $Grad$ represents gradient magnitude.

The gradient and well-exposedness are combined as:
\begin{equation}
    \gamma_{un\exp} =\frac{ wellexpo_{un \exp} \times Grad_{un \exp}}{\sqrt{( wellexpo_{un \exp})^{2}+ Grad_{un \exp}^{2}}} 
\end{equation} 
where $wellexpo$ indicates how close a pixel intensity is to the optimal mid-range exposure.

Similarly, the well-exposedness and variance are combined as:
\begin{equation}
     \gamma_{un\exp} =\frac{\sigma^2_{un \exp} \times wellexpo_{un \exp}}{\sqrt{(\sigma_{un \exp}^{2})^{2}+ wellexpo_{un \exp}^{2}}} 
\end{equation}

All $\gamma$ parameters above are defined for the under-exposed image; replacing the under-exposed image with the over-exposed one yields $\gamma$ values for that exposure. Figure~4.1 illustrates the visual impact of these different definitions on sample inputs.

\begin{figure}[!ht]
  \centering
  \subfloat[Un-exp  image]{\includegraphics[height=2cm,width=3cm]{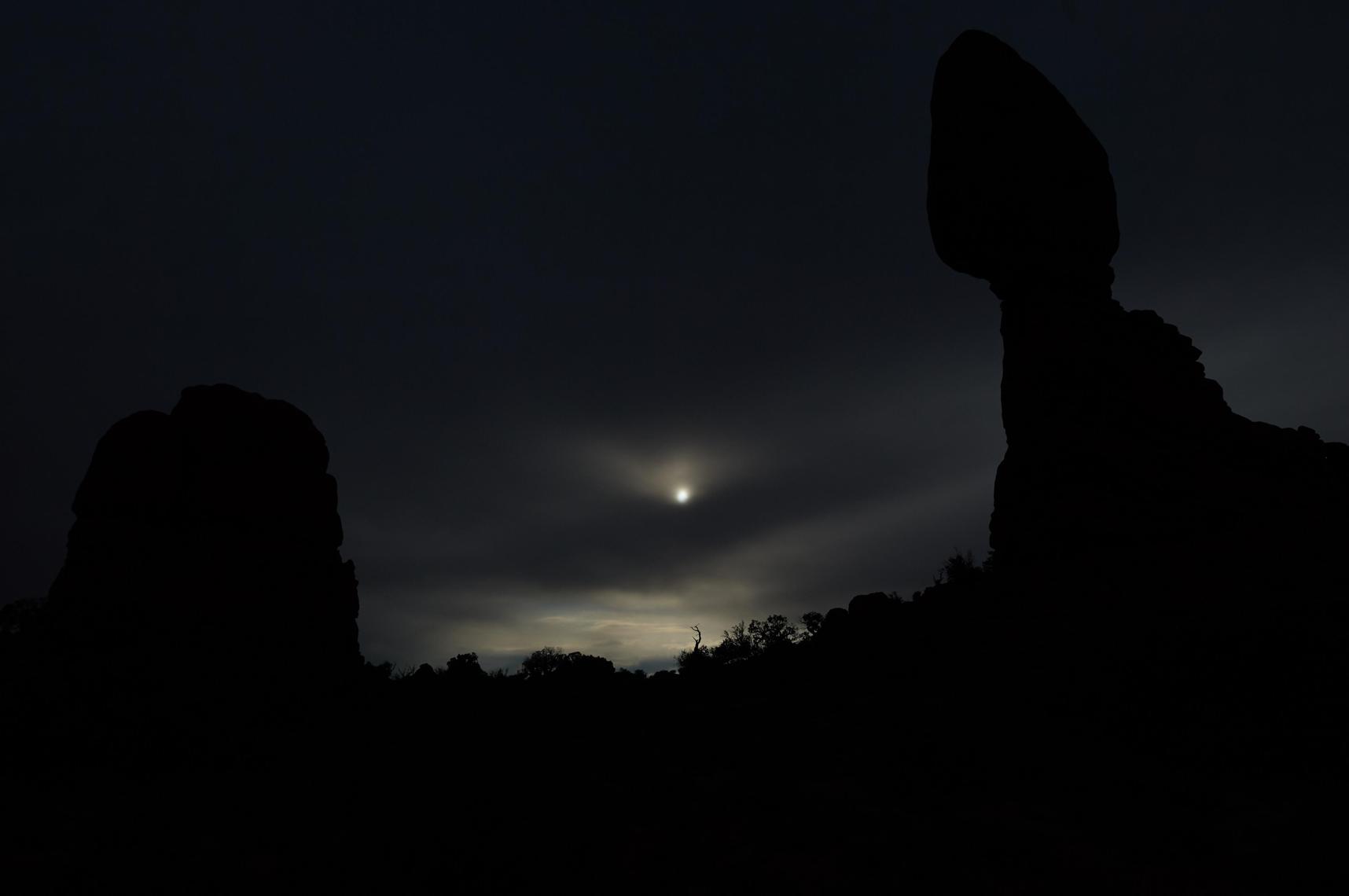}}\qquad
  \subfloat[Ov-exp image]{\includegraphics[height=2cm,width=3cm]{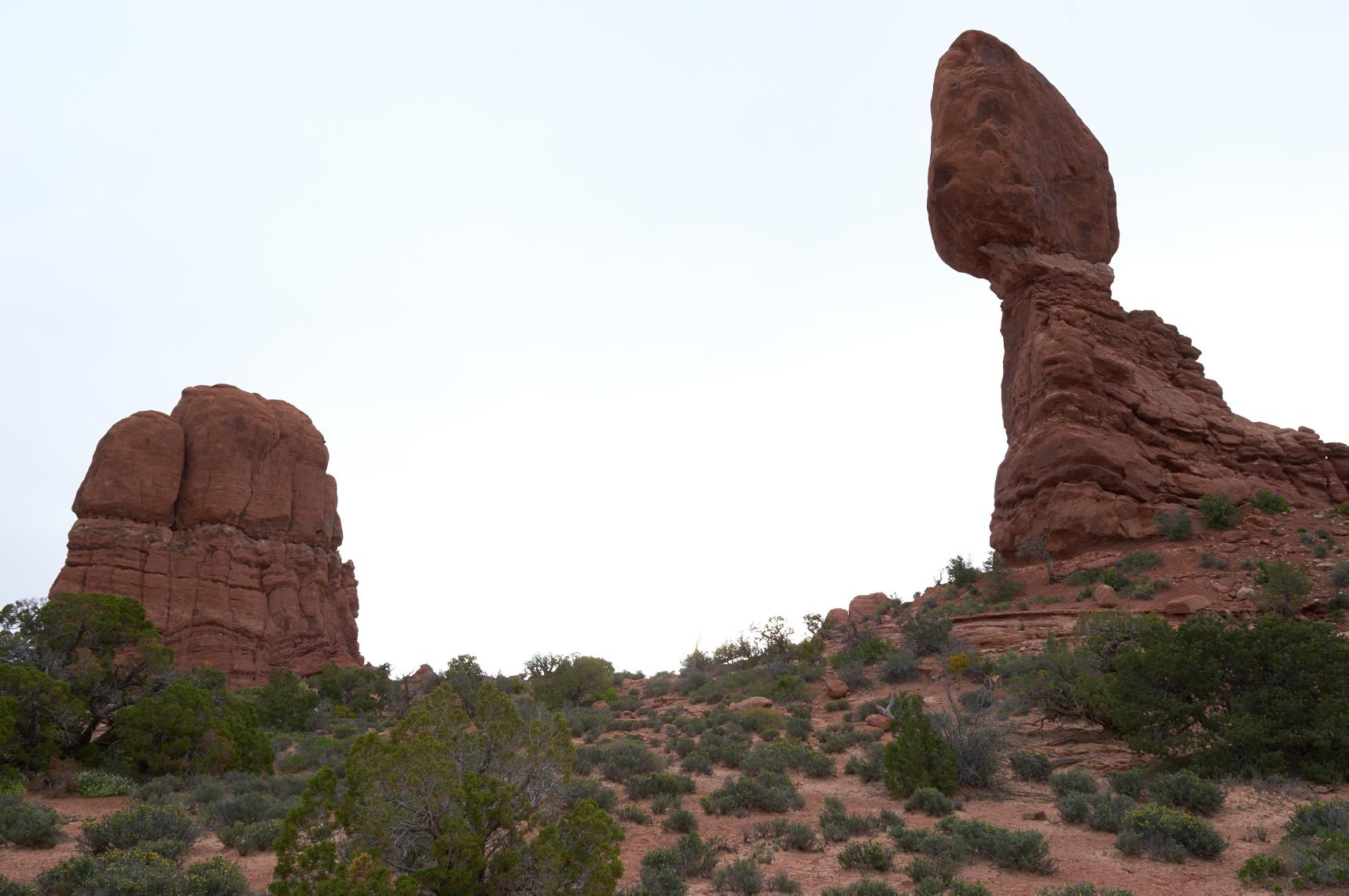}}\qquad
  \subfloat[Variance of Un-exp image]{\includegraphics[height=2cm,width=3cm]{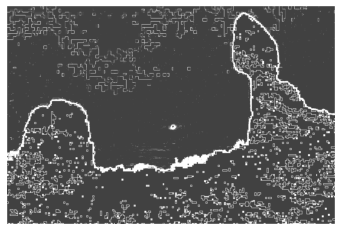}}\qquad
  \subfloat[Variance of Ov-exp image]{\includegraphics[height=2cm,width=3cm]{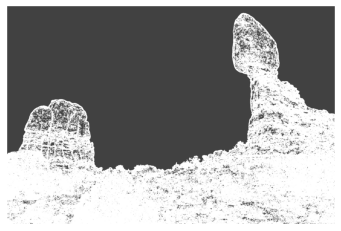}}\qquad
  \subfloat[Gradient of Un-exp image]{\includegraphics[height=2cm,width=3cm]{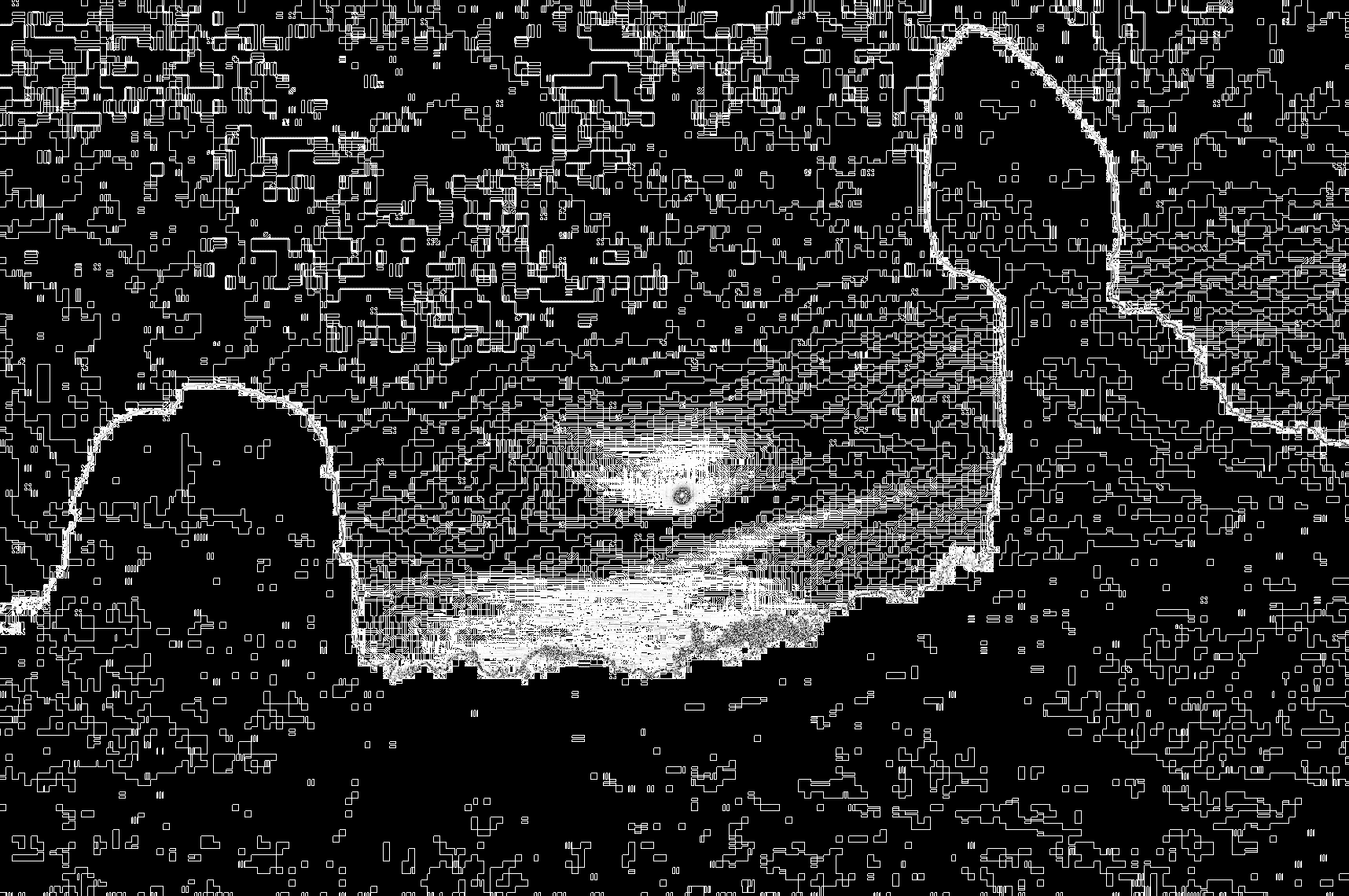}}\qquad
  \subfloat[Gradient of Ov-exp image]{\includegraphics[height=2cm,width=3cm]{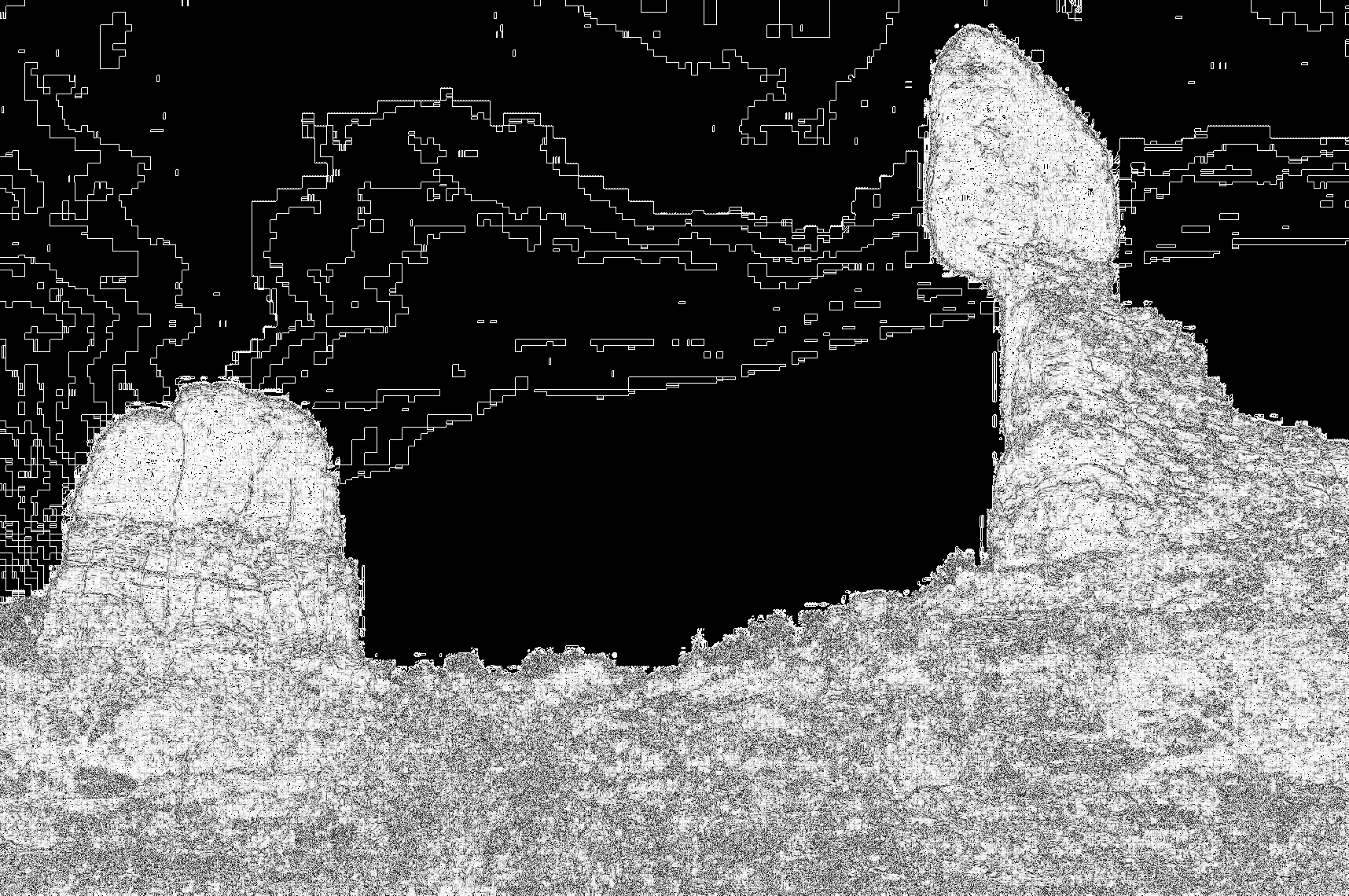}}\qquad
  \subfloat[Well-exposedness of Un-exp image]{\includegraphics[height=2cm,width=3cm]{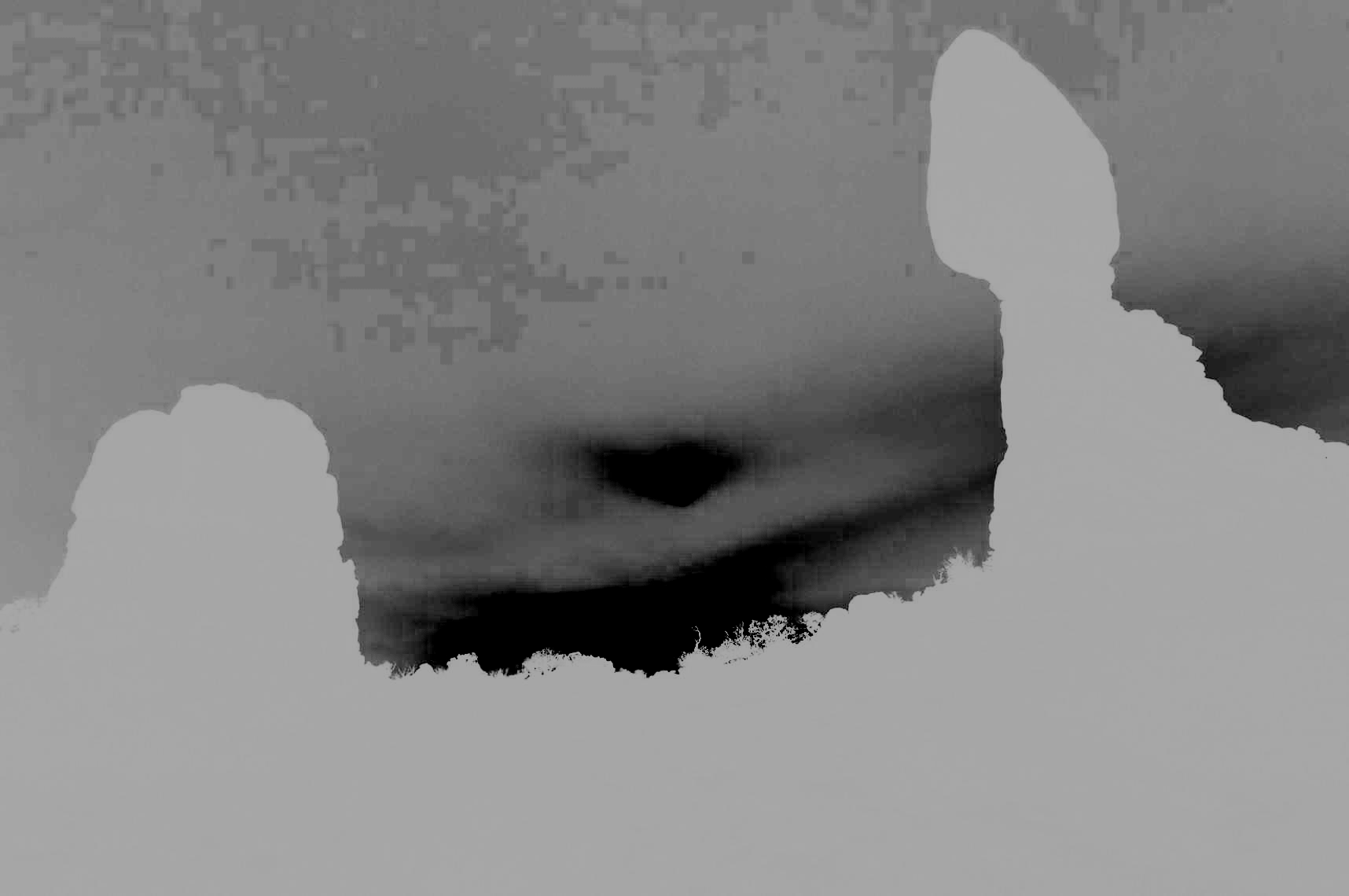}}\qquad
  \subfloat[Well-exposedness of Ov-exp image]{\includegraphics[height=2cm,width=3cm]{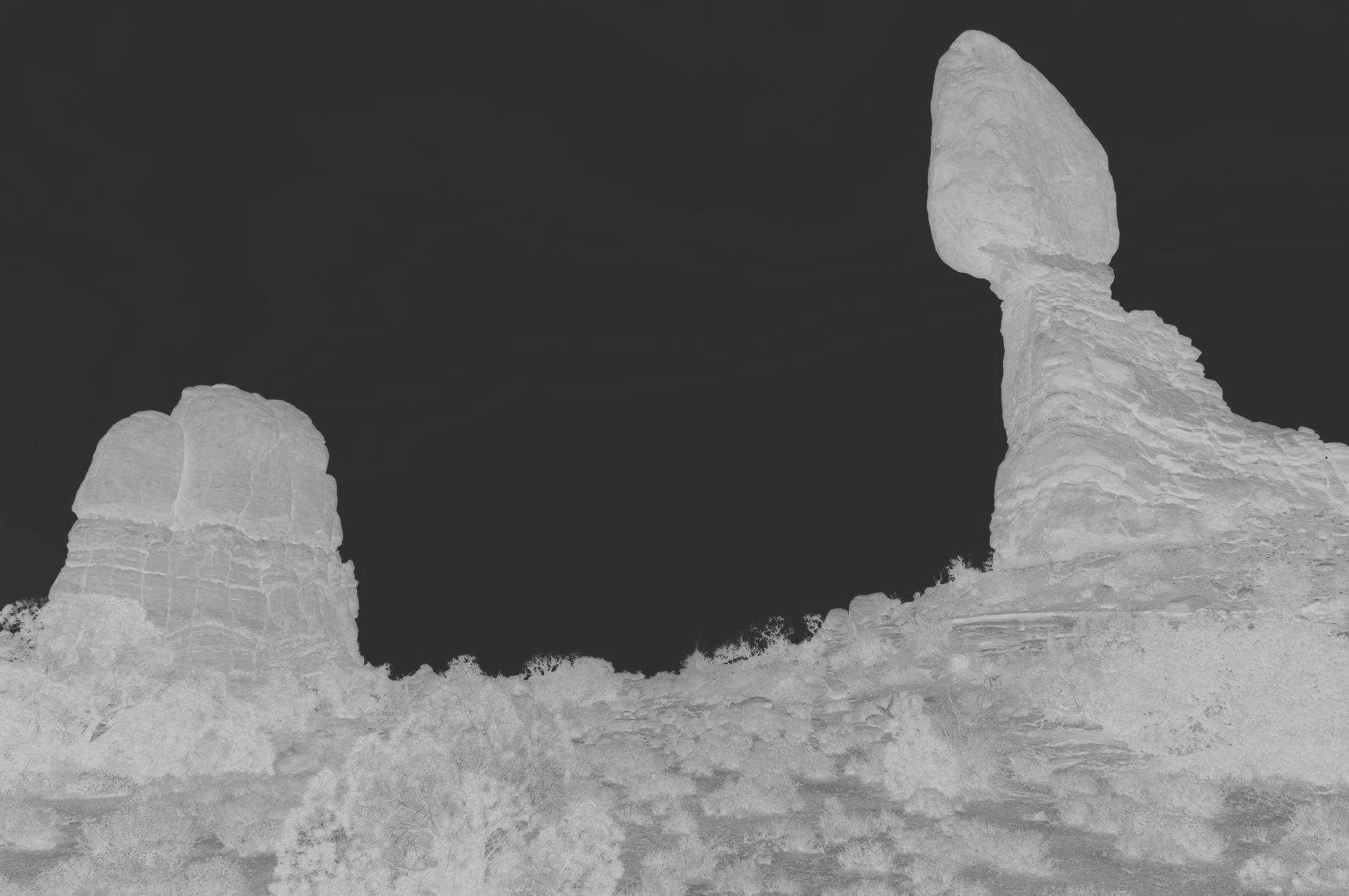}}\qquad
  \subfloat[Var + Grad (Un-exp)]{\includegraphics[height=2cm,width=3cm]{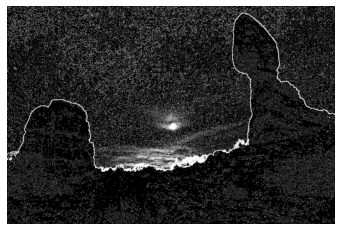}}\qquad
  \subfloat[Var + Grad (Ov-exp)]{\includegraphics[height=2cm,width=3cm]{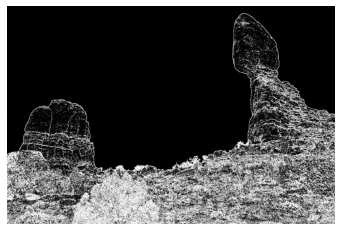}}\qquad
  \subfloat[Var + Well-exposedness (Un-exp)]{\includegraphics[height=2cm,width=3cm]{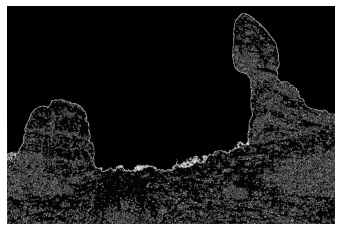}}\qquad
  \subfloat[Var + Well-exposedness (Ov-exp)]{\includegraphics[height=2cm,width=3cm]{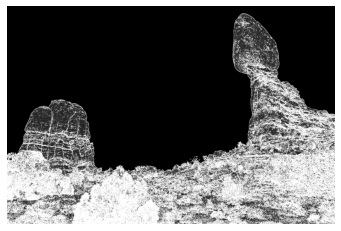}}\qquad
  \subfloat[Grad + Well-exposedness (Un-exp)]{\includegraphics[height=2cm,width=3cm]{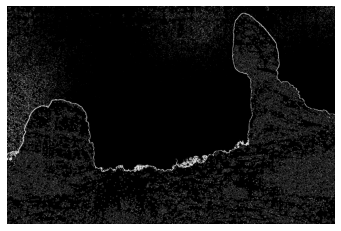}}\qquad
  \subfloat[Grad + Well-exposedness (Ov-exp)]{\includegraphics[height=2cm,width=3cm]{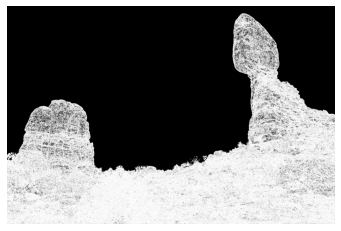}}\qquad
 
  \caption{Gamma visualization on under and over-exposed images. The combination of variance and gradient extracts fine structural details, while other combinations yield smoother but less textured regions.}
\end{figure}

As observed in Figure~4.1(c)–(h), each $\gamma$ component contributes differently. Variance-based maps highlight high-frequency texture regions but can lose subtle contrast variations. Gradient-based $\gamma$ captures edge and boundary information with fine detail, clearly visible in Figures~4.1(e) and~4.1(f). Well-exposedness yields smoother responses, emphasizing balanced luminance over texture sharpness. When parameters are combined, variance–gradient (Figures~4.1(i)–(j)) produces the most informative $\gamma$, integrating both global and local detail cues. In contrast, variance well-exposedness and gradient well-exposedness combinations generate smoother transitions, suitable for uniform luminance regions. These observations confirm that the hybrid variance gradient formulation provides a more balanced weighting strategy, leading to superior HDR fusion results in the loss optimization process.

\section{Experiments and Results}
\subsection{Dataset and Training}
We used dataset from \cite{dataset}, where each image is divided into $250\times250$ patches. Training uses the Adam optimizer \cite{adam} with a batch size of 64, initial learning rate $10^{-4}$, decayed exponentially by 0.99 per epoch.

\begin{table}[ht!]
\centering
\begin{tabular} { | m {4em}| m {4em} | m {4em} | m {5em} | m {4em} | m {6em} | m {6em} |  }
 \hline
 \bf{Test images} & \bf{Variance} & \bf{Gradient} &\bf{Well-exposed ness}  &\bf{Gradient and Well-exposed ness}  &\bf{Variance and Gradient}\\
 
 \hline
  Resort  & 0.9288  & 0.9742  & 0.7582 & 0.9794 & 0.9872  \\
\hline
 Garden  & 0.9575  & 0.9868  & 0.8004  & 0.9894 & 0.9962 \\
\hline
 Buildings  & 0.9740  & 0.9871  & 0.8950 & 0.9804  & 0.9891 \\
\hline
 Doors  & 0.9673  & 0.9902  & 0.8323   & 0.9872 & 0.9970\\
\hline
 Class  & 0.9635  & 0.9902  & 0.8783   & 0.9886 & 0.9983\\
\hline
 Table  & 0.9528  & 0.9805  & 0.8940   & 0.9803 & 0.9878\\
\hline
 University  & 0.9369  & 0.9730  & 0.7177  & 0.9816 & 0.9841\\
\hline
 Statue  & 0.9723  & 0.9903  & 0.8380  & 0.9939  & 0.9976\\

\hline
 Average  & \textbf{0.9566}  & \textbf{0.9840}  & \textbf{0.8267}  & \textbf{0.9851} & \textbf{0.9921} \\
\hline
\end{tabular}
\caption{\label{tab:table-name} MEF-SSIM scores comparison for different proposed methods}
\end{table}
\vspace{10cm}
\begin{figure}[ht!]
  \centering
  \subfloat[Un-exp image]{\includegraphics[height=3cm,width=4cm]{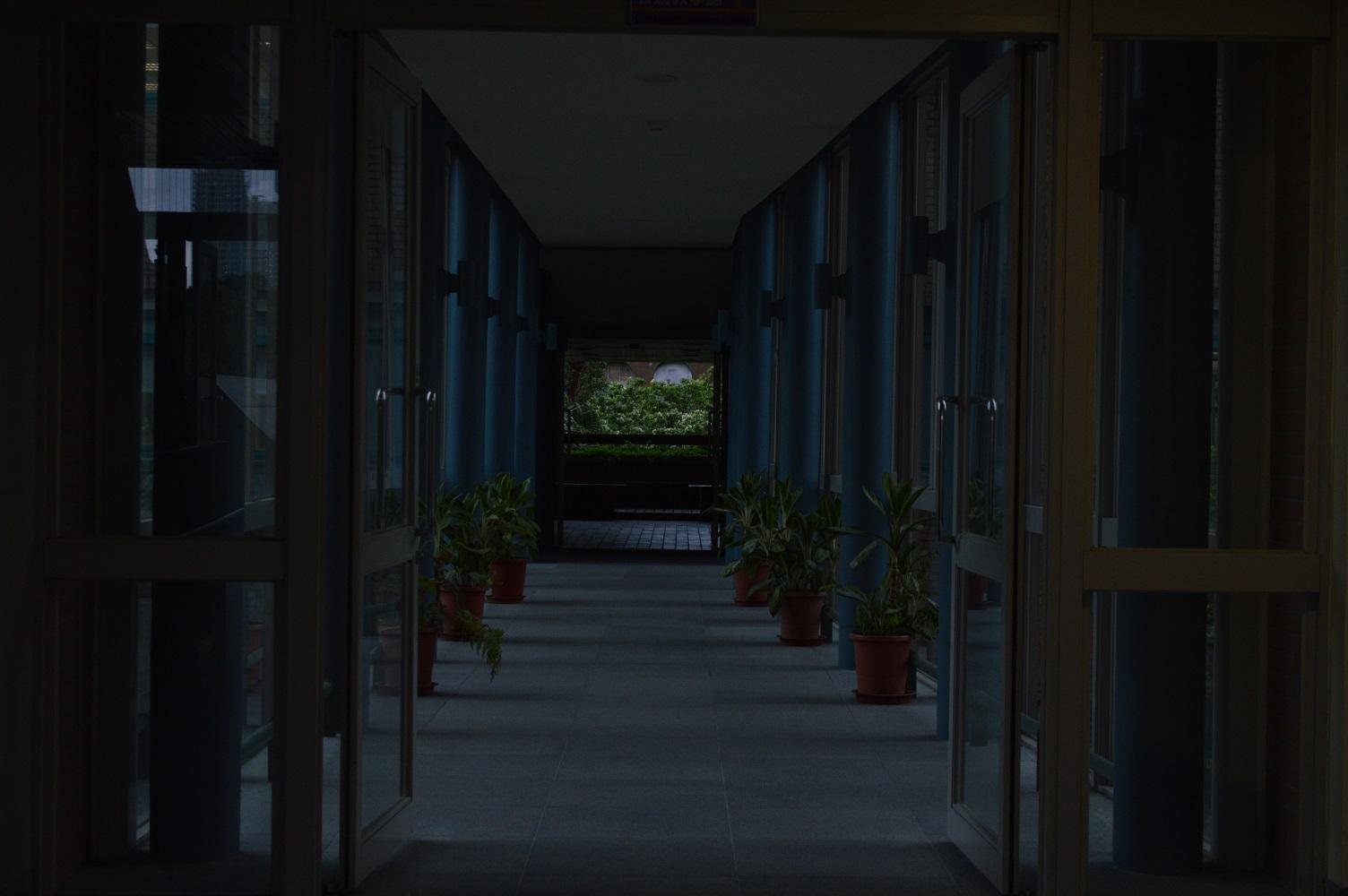}}\qquad
  \subfloat[Ov-exp image]{\includegraphics[height=3cm,width=4cm]{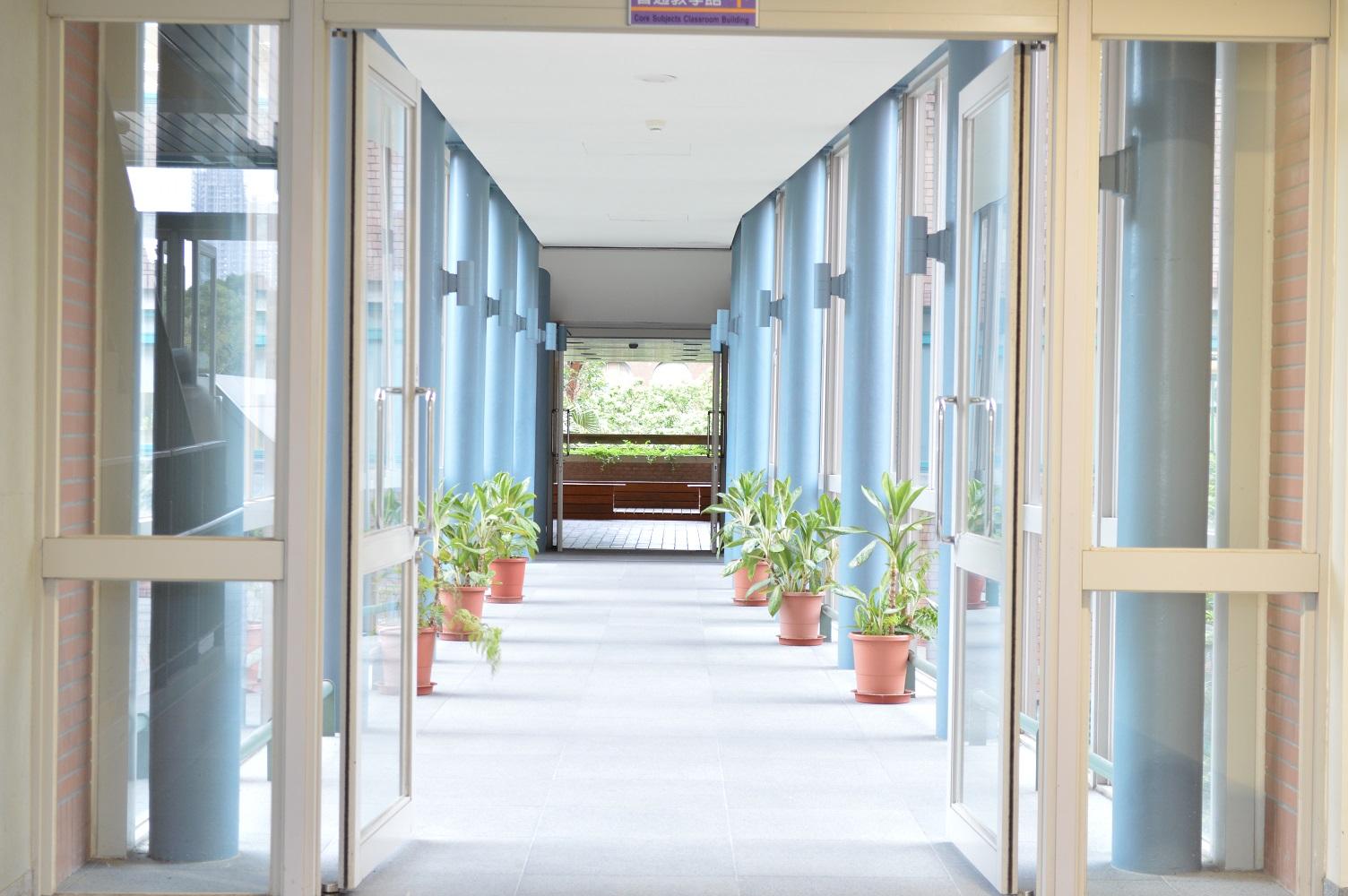}}\qquad
  \subfloat[HDR for variance as $\gamma$]{\includegraphics[height=3cm,width=4cm]{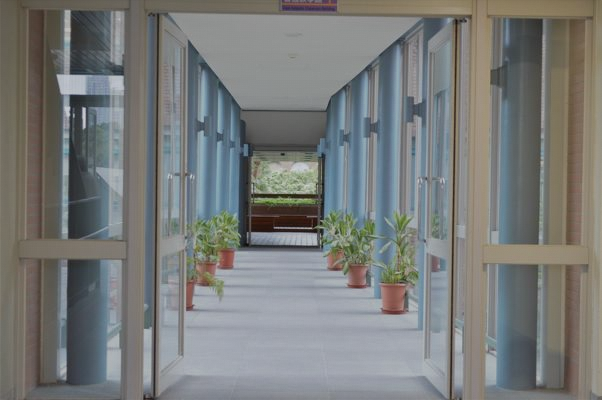}}\qquad
  \subfloat[HDR for gradient as $\gamma$]{\includegraphics[height=3cm,width=4cm]{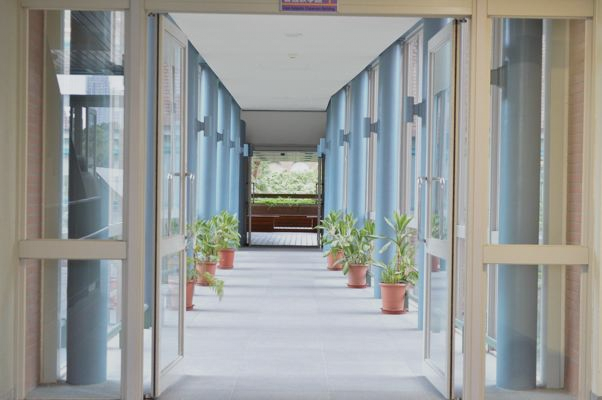}}\qquad
  \subfloat[HDR for well-exposedness as $\gamma$]{\includegraphics[height=3cm,width=4cm]{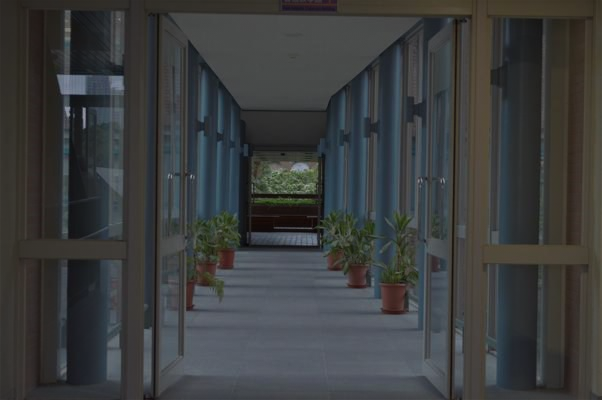}}\qquad
  \subfloat[HDR for variance + gradient as $\gamma$]{\includegraphics[height=3cm,width=4cm]{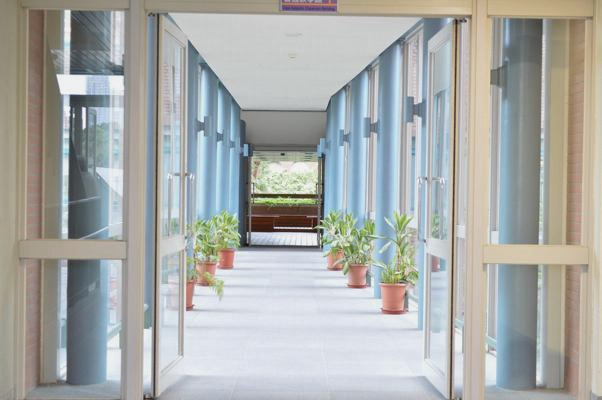}}\qquad
  \subfloat[HDR for gradient + well-exposedness as $\gamma$]{\includegraphics[height=3cm,width=4cm]{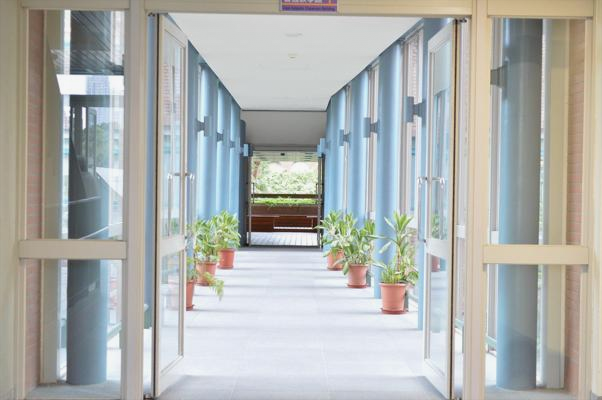}}\qquad
  \subfloat[HDR for variance + well-exposedness as $\gamma$]{\includegraphics[height=3cm,width=4cm]{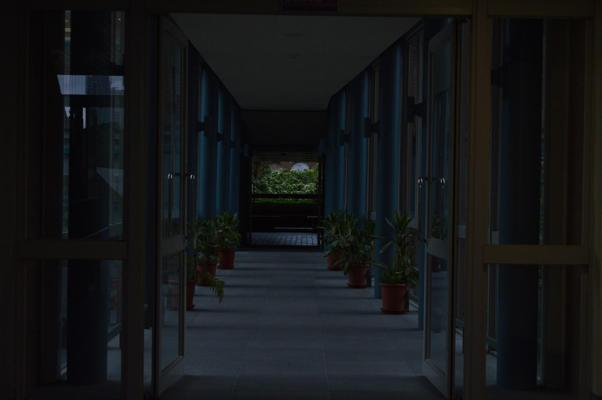}}
  \caption{HDR reconstructions for different $\gamma$ definitions on a test image. The hybrid variance–gradient formulation yields the most balanced and detailed reconstruction.}
\end{figure}

The experimental results in Figure~5 clearly show that the definition of $\gamma$ has a major influence on HDR reconstruction quality. Using \textit{variance} as $\gamma$ emphasizes global contrast but may lose fine textures. In contrast, the \textit{gradient}-based $\gamma$ highlights local edges and structure, capturing fine details in well-exposed regions. The \textit{well-exposedness} parameter produces smoother outputs but lacks high-frequency detail, making it suitable for tone consistency rather than structural sharpness.

Hybrid $\gamma$ formulations provide the best trade-off. Specifically, combining \textit{variance and gradient} produces the highest MEF-SSIM scores across all test sets (average 0.9921), yielding images that maintain texture details, color consistency, and balanced luminance across exposures. These observations confirm that the proposed weighted SSIM loss effectively guides the model to adaptively fuse under and over-exposed inputs, achieving high-quality HDR outputs without the need for ground-truth HDR supervision.

\section{Conclusion}

In this work, the reconstruction of High Dynamic Range (HDR) images was studied using both traditional multi-exposure fusion (MEF) techniques and deep learning-based approaches. Initially, the fundamental process of image capture using a digital camera was discussed, highlighting how exposure settings influence image intensity, background, and possible blur. The motivation for HDR imaging, its advantages over Low Dynamic Range (LDR) imaging, and the various techniques for generating HDR images were also presented. 

The MEF-based approach was first implemented to combine multiple differently exposed images. Two adaptive weighting functions were employed to fuse under and over-exposed inputs, improving luminance balance and structural preservation. The performance of this approach was evaluated using the Multi-Exposure Fusion Structural Similarity Index (MEF-SSIM), which served as a perceptual quality metric to assess reconstruction fidelity. While the MEF method achieved satisfactory results, its performance varied across different test scenes due to handcrafted weighting limitations.

To overcome these challenges, an unsupervised deep learning framework was developed for HDR image reconstruction. Unlike supervised methods, this approach does not require ground truth HDR images, making it more practical for real-world scenarios where such datasets are unavailable. The model learns to generate per-pixel weight maps through a customized loss function based on a modified Structural Similarity Index Measure (SSIM). This weighted SSIM loss enhances the network’s ability to capture complementary details from both underexposed and overexposed inputs.

Further, a refined loss formulation was proposed by varying the $\gamma$ parameter using different perceptual attributes variance, gradient, and well-exposedness and their combinations. Experimental results demonstrated that the hybrid variance gradient configuration achieved the highest MEF-SSIM scores, indicating superior preservation of fine details, contrast, and luminance consistency. 

Overall, the proposed unsupervised CNN model with the customized weighted SSIM loss effectively reconstructs high-quality HDR images from multiple exposures without requiring reference HDR data. The study confirms that adaptive weighting guided by structural and statistical cues leads to more visually accurate and perceptually consistent HDR reconstructions for static scenes.

\bibliographystyle{unsrt}  
\bibliography{references}     

\end{document}